\documentclass[hyper,letterpaper]{JHEP3} 

\JHEPspecialurl{http://jhep.sissa.it/JOURNAL/JHEP3.tar.gz}

\usepackage{epsfig,multicol,amssymb}

\newcommand\fverb{\setbox\pippobox=\hbox\bgroup\verb}
\newcommand\fverbdo{\egroup\medskip\noindent%
            \fbox{\unhbox\pippobox}\ }
\newcommand\fverbit{\egroup\item[\fbox{\unhbox\pippobox}]}
\newbox\pippobox

\title{Dynamically Warped Theory Space and Collective Supersymmetry Breaking}

\author{Christopher D. Carone, Joshua Erlich, and Brian Glover \\
        Particle Theory Group, Department of Physics,
College of William and Mary, Williamsburg, VA 23187-8795 \\
    E-mail: \email{carone@physics.wm.edu}, \email{erlich@physics.wm.edu},
    \email{baglov@wm.edu}}
\received{blah}        
\revised{blah}
\accepted{blah}        

\preprint{WM-05-113}

\abstract{We study deconstructed gauge theories in which a warp factor emerges
dynamically and naturally.  We present nonsupersymmetric models in which the
potential for the link fields has translational invariance, broken only by boundary
effects that trigger an exponential profile of vacuum expectation values.
The spectrum of physical states deviates exponentially from that of the
continuum for large masses; we discuss the effects of such exponential towers
on gauge coupling unification.  We also present a supersymmetric example in
which a warp factor is driven by Fayet-Iliopoulos terms. The model is peculiar in
that it possesses a global supersymmetry that remains unbroken despite nonvanishing
D-terms.  Inclusion of gravity and/or additional messenger fields leads to the collective
breaking of supersymmetry and to unusual phenomenology.
}


\begin{document}


\section{Introduction}\label{sec:intro}
The possibility exists that there are spatial dimensions beyond the
three that we are able to perceive. However, gauge theories in extra
dimensions are not renormalizable and are to be understood as
effective theories.  An example of an ultraviolet completion of extra-dimensional
gauge theories is provided by deconstruction~\cite{Nima,Fermilab}. In this approach, one
arranges the action of a four-dimensional gauge theory so that at low energies it reproduces
the action of a five-dimensional theory that is latticized in one spatial dimension.
The high-energy theory can be asymptotically free, with fermions condensing at
an intermediate scale to provide the link fields of the latticized theory~\cite{Nima}.
At energies below the scale at which the condensation occurs (which still must
be significantly higher than the inverse size of the extra dimension) the
theory is effectively described by a ``moose'' theory in which scalar link
fields obtain vacuum expectation values (vevs)~\cite{Fermilab}.  The spectrum
of fields in the spontaneously-broken theory is designed to mimic the
Kaluza-Klein (KK) towers of the higher-dimensional theory.

Deconstructed extra dimensions have proven useful for building the
features of higher-dimensional phenomenology into four-dimensional models.
Examples include the little Higgs models of electroweak symmetry
breaking~\cite{little-higgs}, extra-dimensional grand unified theories~\cite{guts}
and models of supersymmetry (SUSY) breaking~\cite{CEGK,CKSS,twisted-susy,pokorski,dienes,japanese}.

Deconstruction has also provided a new handle on nonperturbative effects in
higher-dimensional supersymmetric theories and string
compactifications~\cite{exact-results,0-2}.  The novel latticizations which
arise in studies of deconstruction have led to new approaches in latticizing
chiral gauge theories~\cite{lattice} and supersymmetric theories~\cite{susy-lattice}.
The deconstruction of theories with higher-dimensional gravity has not been
completely successful, but has provided insight into the scales at which
gravitational modes in a latticized theory become strongly coupled~\cite{gravity}.
In addition, deconstruction of warped extra dimensions~\cite{Sfetsos,RSW} has allowed for
an explicit realization of the holographic renormalization group~\cite{holo-RG}
and the transition between logarithmic and power-law running of couplings
as a function of energy~\cite{power-law}.

Most deconstructed extra-dimensional models are fine tuned in the sense that
the gauge couplings at each lattice site and the vevs
of the link fields must be fixed precisely in order to reproduce the dynamics of an
extra dimension.  We will study some theories in which a warped theory space is
generated dynamically, without a significant fine-tuning of parameters. To this end,
we impose an approximate hopping symmetry in the link field potential, which would render
our theories invariant under translations if our moose were infinitely long.   We allow the
form of the hopping potential to vary at the ends of our finite moose, as a way of taking
into account effects that could reasonably occur at the boundaries in the continuum theory.
With these assumptions, we will see that an exponential profile of link field vevs can
occur naturally for a wide range of model parameters, precisely what one needs to
deconstruct models with bulk fields in Anti-de Sitter (AdS) space.

In contrast to the spectra obtained in the continuum theory, however, we find towers of
gauge and link field states whose masses grow exponentially with KK number.
This suggests a concrete way to distinguish the four-dimensional models that we study
from those that genuinely live in a warped extra dimension.
If the standard model exists in the bulk of the deconstructed extra dimension, we will see
that unification occurs at least as well as in the nonsupersymmetric standard model, but at
an accelerated rate due to the Kaluza-Klein modes.

We then turn to supersymmetric moose models that dynamically generate warp factors.
We focus on a U(1)$^n$ theory in which a Fayet-Iliopoulos term for each gauge factor
forces vevs for neighboring link fields to vary monotonically along the
lattice, except at the boundary.
What is intriguing about this model is that each U(1) factor has a nonvanishing
D term, yet the low-energy theory remains globally supersymmetric, as we will see by
studying the spectra of the gauge and link fields.  We explain how this unusual
circumstance is possible, and then add mediating fields and gravity, leading to
nonsupersymmetric spectra.  The collective SUSY breaking in the extended theory
shares some features with Scherk-Schwarz SUSY breaking~\cite{Scherk--Schwarz} and twisted
theory space~\cite{twisted-susy}, but is also different from those types of models in
several important ways.  In the absence of the mediating fields, global SUSY remains
unbroken, as opposed to the usual situation in which mediating fields have a tendency
to restore SUSY.  Also, SUSY breaking here is of the ``supersoft" type~\cite{supersoft}
because there are no fields which obtain F-term vevs.  However, the suppression
of the SUSY breaking scale with respect to the vacuum energy yields a gravitino that is
heavy in these models.

Our paper is organized as follows. In Section~\ref{sec:frame} we review the framework of
deconstructed extra dimensions and theory space.  In Section~\ref{sec:nonsusy} we study
dynamically generated nonsupersymmetric warped extra dimensions.  We study gauge coupling
running in these deconstructed theories and compare with the continuum theory and with the
Standard Model.  In Section~\ref{sec:susy} we study the dynamically deconstructed
supersymmetric U(1) theory and its unusual SUSY breaking phenomenology. We conclude in
Section~\ref{sec:conc}.

\section{Framework}\label{sec:frame}
In this section we review the deconstruction of a warped extra dimension.
It is important to note that the gauge theories we eventually study are
more correctly described as models in ``theory space", {\em i.e.}, the space of
four-dimensional theories that can be described conveniently by moose diagrams.  For
particular values of the link field vevs and gauge couplings, the theory space will
coincide at low energies with a latticized extra dimension.  It will be of particular
interest to us when this correspondence reproduces the spectrum and interactions of a
gauge theory in AdS space, at least in the continuum limit.

\EPSFIGURE{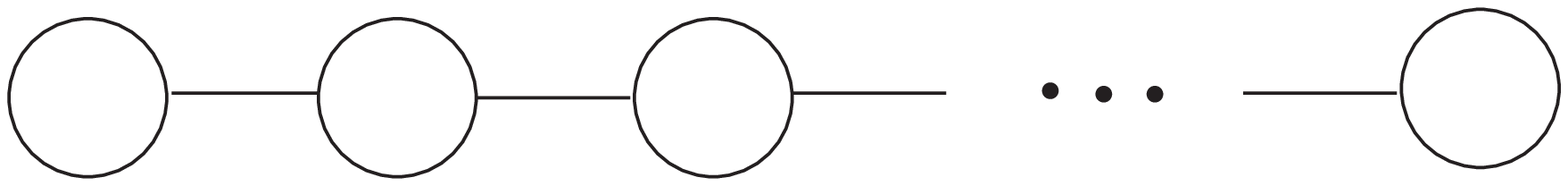,width=4in}{A linear moose. \label{fig:moose}}
Consider an $n$-site moose diagram with $n$ SU(N) gauge groups connected by $n-1$ link
fields $\phi_i$, as shown in Fig.~\ref{fig:moose}.  The link fields transform in the
bifundamental representation under neighboring groups. The action of our moose diagram is
\begin{eqnarray}
    S = \int d^4x \sum_i\left[-\frac{1}{2}\hbox{Tr}\left[F^{\mu\nu}_i
    F^i_{\mu\nu}\right]+\hbox{Tr}|D_\mu\phi_i|^2\right] \,
    \label{eq:mac}
\end{eqnarray}
where $F^{\mu\nu}_i$ is the non-Abelian field strength for the $i^{th}$ group,
and $\phi_i$ is the link field connecting the $i^{th}$ and $(i+1)^{th}$ site. (We leave
it implicit that the sum runs up to $n$ in the first term and $n-1$ in the second term.)
The covariant derivative of $\phi_i$ is given by, \begin{equation}
D_\mu \phi_i=\partial_\mu\phi_i+iq\,A_\mu^i\phi_i-iq\,\phi_iA_\mu^{i+1}.
\end{equation}
Note that
$F_{\mu\nu}\equiv F_{\mu\nu}^a T^a$, where $T^a$ is a group generator, satisfying
$\mbox{Tr}(T^a T^b)=\delta^{ab}/2$. (In the Abelian case we will take
the generators to be the identity, so that $\mbox{Tr}(T^2)=1$).
To compare the moose theory with the lattice gauge theory action it is
useful to express Eq.~(\ref{eq:mac}) using a nonlinear field redefinition of
the bifundamental fields $\phi_j$:
\begin{eqnarray}
\phi_j = v_j \, U(j,j+1) &\equiv& v_j\,\exp\left[iqa\,A_5^j\right] \nonumber \\
&\approx& v_j\, (1 + i q a A_5^j + \cdots)\,,
\end{eqnarray}
where we have expressed the $j^{th}$ link in terms of a ``comparator'' field
$U$.  The parameter $a$ has units of length, and the exponential in the field
redefinition represents the
Wilson line along the interval $(y_j,y_j+a)$ in the extra dimension.
With this choice, the action
for the link fields may be written to quadratic order in $A_5$ as,
\begin{eqnarray}
S &=& \int d^4 x \sum_i q^2v_i^2 a^2\mbox{Tr} |\partial_\mu
    A_5^i - \frac{(A^{i+1}_\mu - A_\mu^{i})}{a}+iq\,(A_\mu^iA_5^i-A_5^{i}
A_\mu^{i+1})|^2 \nonumber \\
&=&\int d^4 x \sum_i q^2v_i^2 a^2\mbox{Tr} |\partial_\mu
    A_5^i - \frac{(A^{i+1}_\mu - A_\mu^{i})}{a}+iq\,[A_\mu^i,A_5^i]
-iqA_5^i(A_\mu^{i+1}-A_\mu^i)|^2
\end{eqnarray}
We can compare this to the action of a warped five-dimensional
theory,
\begin{equation}
    S = \int \,d^4x \,dy \sqrt{g} \, \left(
        -\frac{1}{2} \mbox{Tr}[F^{\mu\nu}F_{\mu\nu}]-
        \mbox{Tr}[F^{\mu5}F_{\mu5}]\right)\,,
\end{equation}
where the metric is of the form
\begin{equation}
ds^2 = e^{-f(y)}dx^2+dy^2 \,. \label{eq:metric}
\end{equation}
We now discretize this action on a lattice with $n$ sites and spacing $a$, so that the
length of the extra dimension in these coordinates is $R=na$.  This requires that we make
the substitutions
\begin{equation}
F^{\mu\nu}(x^\rho,y) \rightarrow F^{\mu\nu}_i(x^\rho)\,,\,\,\,\,\,\,f(y)\rightarrow f_i \,,
\end{equation}
\begin{equation}
\int dy \rightarrow \sum_i a \,\,\,,
\end{equation}
\begin{equation}
\partial_5 A^{\mu} \rightarrow \frac{A_{i+1}^{\mu} - A_{i}^{\mu}}{a} \,\,\,,
\end{equation}
so that our five-dimensional action becomes
\begin{equation}
S = \sum_i a \int d^4x\, \left[ -\frac{1}{2}\mbox{Tr}(F_i^{\mu\nu})^2
+ e^{- f_i}\,\mbox{Tr}\left(\partial_\mu A_i^5 -
\frac{A^{i+1}_\mu-A^{i}_\mu}{a}+iq\,[A_\mu^i,A_5^i]\right)^2 \right] \,\,\,.
\end{equation}
Thus we recover the action of the moose model, up to terms of higher order
in the lattice spacing $a$, provided we identify
\begin{equation}
v_i = \frac{e^{-f_i/2}}{qa} \,\,\,.
\end{equation}

The exercise above demonstrates how the geometry of an extra dimension may be encoded in
the profile of link field vevs in the four-dimensional theory.  For AdS space, one has
$f_j = 2 k a j$, where $k$ is the AdS curvature, and one finds that $v_{j+1} = \exp(-ka)\, v_j$.
In the next section, we present nonsupersymmetric models in which such a profile of vevs is
generated dynamically.   In these models, the potential for the link fields will have the form
\begin{equation}
V = V_b\,(\phi_1)+ \sum_{i=1}^{n-1} V_i\,(\phi_i,\phi_{i+1})\,,
\label{eq:genform}
\end{equation}
with $\phi_n\equiv0$, corresponding to a Dirichlet boundary condition. The second term in
Eq.~(\ref{eq:genform}) is invariant under translations $i\rightarrow i+1$ along the moose,
except at the boundaries $i=1$ and $i=n-1$. We may choose $V_b(\phi_1)$ so that
the explicit breaking of this translation invariance triggers a monotonically varying
profile of link vevs.  We show in the next section that generic forms exist for the hopping
terms $V_i$ that provide for local minima with the desired properties, without requiring an
unnatural choice of parameters.

The deconstruction of warped supersymmetric theories is similar to the example just considered,
except that each site corresponds to an ${\cal N}=1$ vector multiplet, and each link to a
chiral multiplet.  Aside from the vector multiplet that includes the zero mode gauge field,
the remaining $n-1$ massive multiplets include both a vector and chiral superfield, each
transforming as an adjoint under the unbroken diagonal subgroup. This is consistent with our
expectation that the KK levels arising from an underlying 5D theory should form ${\cal N}=2$
SUSY multiplets. Each bi-fundamental link field also contains a singlet under the unbroken
gauge group, but these states are generally assumed to have no impact on the low-energy
phenomenology.  In our explicit treatment of the link potential, the spectra of all physical
states will be specified in our models.

\section{Non-supersymmetric Warped Theory Space}\label{sec:nonsusy}

In this section, we consider non-supersymmetric examples of the
class of theories described in Section~\ref{sec:frame}. These
theories dynamically generate a warp factor that becomes
exponential in the limit that the number of sites $n$ is taken
large.  We determine the spectrum of link and gauge fields that
appear in such models later in this section.

\subsection{Non-supersymmetric Models}

Let us begin by considering nonsupersymmetric U(1)$^n$ gauge
theories, with gauge couplings $g_i=g$ and $n-1$ link fields with
charges
\begin{equation}
\phi_i \sim (+1,-1) \,\,\, ,
\end{equation}
under the $i^{th}$ and $(i+1)^{th}$ group factors, respectively.
We wish to find potentials of the form (\ref{eq:genform}) that will
generate a warp factor.  As a warm up, consider the following fine-tuned
example,
\begin{equation}
V_b = (a_1-m^2)^2 \,\,\, ,
\label{eq:model1vb}
\end{equation}
and
\begin{equation}
V_i = (\lambda \, a_i-a_{i+1})^2 \,\,\, .
\label{eq:model1v}
\end{equation}
where $a_i = \phi_i^\dagger\phi_i$, for $i=1\dots n-1$, and $a_n=0$. This
model is fine-tuned in the sense that the terms appearing in the potential
are not the most general set allowed by the symmetries of the theory.  We discuss more
general examples afterwards.  The advantage of the model defined by
Eqs.~(\ref{eq:model1vb}) and (\ref{eq:model1v}) is that it allows us
to extract a number of useful results without resorting to numerical
analysis.

Since we are looking for solutions in which all the $\phi$ develop vevs,
we may minimize $V$ with respect to the $a_i$. Excluding the links at the
ends of the moose, the minimization condition for the $\ell^{th}$ link is
simply
\begin{equation}
(1+\lambda^2)\,a_\ell - \lambda\,(a_{\ell+1}+a_{\ell-1})=0 \,\,\,.
\label{eq:minblk}
\end{equation}
Given the translation invariance of the system, it is not hard to
show that a general solution is given by
\begin{equation}
a_j = A\,\exp(Kj) + B\, \exp(-Kj)
\label{eq:gensol}
\end{equation}
where
\begin{equation}
\cosh K  = \frac{1+\lambda^2}{2\lambda} \,\,\,.
\label{eq:transeq}
\end{equation}
The minimization conditions on $a_1$ and $a_{n-1}$ both differ from
Eq.~(\ref{eq:minblk}),
\begin{equation}
(1+\lambda^2)\, a_1 - m^2 - \lambda\, a_2 =0 \,\,\, ,
\end{equation}
\begin{equation}
-\lambda\, a_{n-2} + (1+\lambda^2)\, a_{n-1} =0 \,\,\,.
\end{equation}
and determine the coefficients $A$ and $B$. After some algebra, the general
solution in Eq.~(\ref{eq:gensol}) can be reduced to
\begin{equation}
a_j = \frac{m^2}{\lambda}\left(\frac{1}{1-\lambda^{2n}} \, \lambda^j
+\frac{1}{1-\lambda^{-2n}} \, \lambda^{-j}\right)\,\,\,,
\label{eq:thesol}
\end{equation}
for $j=1\ldots n-1$.

This solution provides the desired warp factor, providing it corresponds
to, at least, a local minimum of the potential.  For our choice $\lambda<1$,
one sees that in the $n\rightarrow \infty$ limit Eq.~(\ref{eq:thesol})
reduces to
\begin{equation}
a_j = m^2 \, \lambda^{j-1} ,
\label{eq:ninfsol}
\end{equation}
since the second term approaches $-\lambda^{2n-j}$, which is vanishing for
all $j$.  For finite $n$, one finds monotonically decreasing link vevs.  For
example, in the case where $n=4$ ({\em i.e.}, three link fields) and
$\lambda=0.3$ one finds
\begin{equation}
v_1 = 1.00\, m, \,\,\,\,\,\,\,\,\,\,
v_2 = 0.55\, m, \,\,\,\,\,\,\,\,\,\,
v_3 = 0.29\, m,
\label{eq:neq4}
\end{equation}
where $v_i = \langle a_i \rangle^{1/2}$.  Note that the U(1)$^n$ gauge
invariance allows us to choose all of the $v_i$ real and positive.

Without working out the mass spectrum of the link fields explicitly,
we may nonetheless show that the extremum of the potential just
described is at least a local minimum.  First we note that in the
$n\rightarrow\infty$ limit, the solution Eq.~(\ref{eq:ninfsol}) corresponds to
$V(\{\langle\phi_i\rangle\})=0$.  However, the potential is clearly positive
definite and there are no flat directions.  We therefore conclude that
our solution in the $n\rightarrow\infty$ limit corresponds to a global minimum.

For $n$ finite, consider the scalar mass squared matrix
\begin{equation}
m^2_{ij} = \left(\frac{\partial^2 V}{\partial \xi_i \xi_j} \right)_{min} =\left(
\frac{\partial^2 V}{\partial a_k \partial a_\ell} \frac{\partial a_k}{\partial \xi_i}
\frac{\partial a_\ell}{\partial \xi_j} \right)_{min} \,\,\, ,
\label{eq:msmat}
\end{equation}
where $\xi_i$ represents the real and imaginary components of the link fields in
the basis
\begin{equation}
\xi = ( \phi_{1}^{im}, \phi_{2}^{im},\ldots,\phi_{n-1}^{im},
\phi_{1}^{re}, \phi_{2}^{re},\ldots,\phi_{n-1}^{re}) \,\,\, .
\end{equation}
For $1\leq i \leq n-1$ or $1\leq j \leq n-1$, the factors $\partial a / \partial \xi$
are vanishing, since the vevs of the $\phi_i$ are purely real.  This implies that there
are $n-1$ zero eigenvalues, corresponding to the goldstone boson degrees of freedom in
the spontaneous breaking of U(1)$^n\rightarrow$U(1).  Only the lower-right $(n-1)\times(n-1)$
block of the Eq.~(\ref{eq:msmat}) is nonvanishing, and is of the form
\begin{equation}
m^2 = V A\,  V^T \,\,\, ,
\end{equation}
where $V$ is a diagonal matrix of vacuum expectation values $V=diag(v_1,v_2,\ldots v_{n-1})$,
and $A$ is a dimensionless matrix of the form
\begin{equation}
A = \left(\begin{array}{cccccc} (1+\lambda^2) & -\lambda  & & & & \\
                                -\lambda & \, (1+\lambda^2) \,\,& -\lambda & & & \\
                                & & & & \ddots & \\
                                & & & & -\lambda & (1+\lambda^2) \end{array}\right) \,\,\,.
\end{equation}
Since $V$ is nonsingular, it follows that the number of positive eigenvalues of $m^2$ and $A$
are the same. Therefore, it is sufficient that we show that $A$ has only positive eigenvalues.
The proof is as follows: For $\lambda=0$, $A$ is the identity matrix, which is clearly positive
definite. As we allow $\lambda$ to vary continuously away from zero, the only way any eigenvalue
can become negative is for there to exist a value of $\lambda$ for which that eigenvalue vanishes
and the determinant of $A$ is zero.  However, one can verify that
\begin{equation}
\det A = 1 + \sum_{j=1}^{n-1} \lambda^{2j} \,\,\, ,
\end{equation}
which is never vanishing.  Thus, all the eigenvalues of $A$, and hence $m^2$, remain positive
for arbitrary $\lambda$.  Our warped solution corresponds to a minimum of the potential.

As we have commented earlier, the potential we have just examined is not the most general
one that we could have written down.  Ignoring the currently fashionable trend of settling
for fine-tuned models, we now turn to more general possibilities.   Assuming renormalizable,
next-to-nearest-neighbor interactions, the most general form for the
$V_i$ in a U(1)$^n$ theory is
\begin{equation}
V_i (\phi_i,\phi_{i+1}) = \hat{m}^2 a_i + \lambda\, a_i^2 + \rho a_i \, a_{i+1} \,\,\, .
\label{eq:mostgenv}
\end{equation}
As before, we assume the boundaries of the moose are special, and include the additional
corrections
\begin{equation}
V_b(\phi_1) = - m_1^2 a_1 + \lambda_1 a_1^2 \,\,\, ,
\label{eq:mostgenvb}
\end{equation}
designed to trigger a vacuum expectation value at one end of the moose.  In this general
parameterization, the four-site example that gave us Eq.~(\ref{eq:neq4}) corresponds to
$m_1^2=2 m^2$, $\lambda_1=0$, $\hat{m}^2=0$, $\lambda=1.09$, and $\rho=-0.6$. By varying these
parameters continuously away from our successful, yet fine-tuned, solution we can
find more general results.  For example, the parameter choice
$\lambda_1=-0.8$, $\hat{m}^2=0.4$, $\lambda=1.0 $, and $\rho=-0.7$ we obtain a local
minimum with
\begin{equation}
v_1 = 1.78\, m_1,  \,\,\,\,\,\,\,\,\,\,
v_2 = 0.98\, m_1, \,\,\,\,\,\,\,\,\,\,
v_3 = 0.37\, m_1 \,\,\, .
\label{eq:moregen}
\end{equation}
For large $n$, where end effects become less important in determining the location of
the minimum, one can find parameter choices that yield viable warped solutions for
arbitrary $n$.  To illustrate this point, we have studied numerically the potential
defined by Eqs.~(\ref{eq:mostgenv}) and (\ref{eq:mostgenvb}) for $n$ ranging from
$30$ to $100$.   The results, which we have verified correspond to minima of the potential,
are shown in Fig.~\ref{fig:3.1}.

\EPSFIGURE{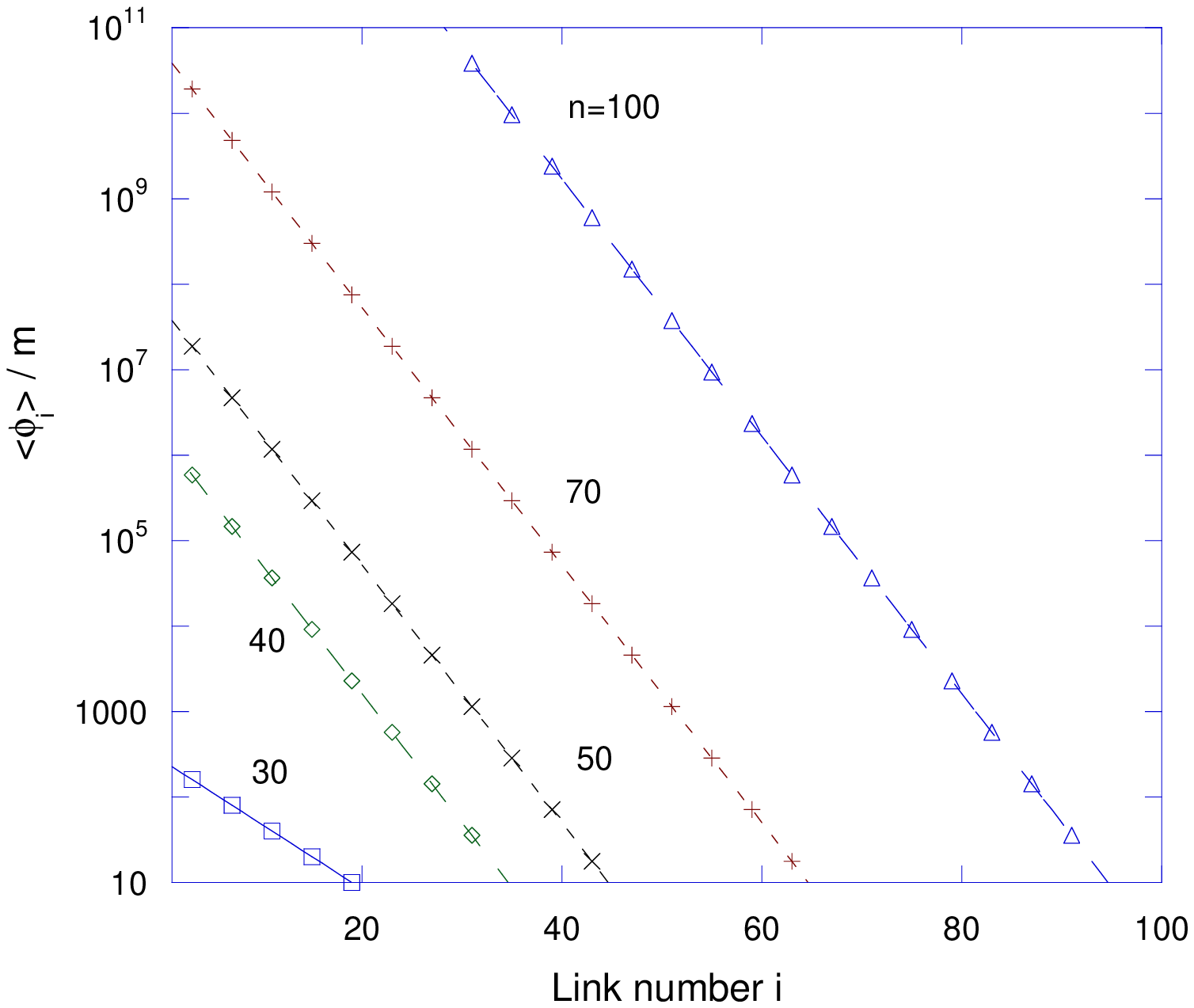,width=4in}{Link field vevs for $n=30$, $40$, $50$, $70$ and $100$.
The solutions shown correspond to the parameter choices $m_1^2=2 m^2$,
$\lambda_1=-0.8$, $\lambda=1$, and $\rho=-0.8$. \label{fig:3.1}}

It is clear by inspection that the vevs have the desired approximate exponential
dependence on link number.  Also note that there is no significant fine-tuning in
the choice of parameters $m_1^2=2 m^2$, $\lambda_1=-0.8$, $\lambda=1$, and $\rho=-0.8$.
While other successful solutions are possible, we will not survey the parameter space.

We instead turn to models that may have somewhat different phenomenological applications,
namely those involving non-Abelian group factors.  We are interested in SU(N)$^n$ moose
that are broken to the diagonal SU(N) while spontaneously generating a warp factor.
Defining the SU(N)$_i \times$SU(N)$_{i+1}$ invariant combinations,
\begin{equation}
a_i = \mbox{Tr}({\phi_i}^\dagger \phi_i) \,,\,\,\,\,
b_i = \mbox{Tr}({\phi_i}^\dagger \phi_i {\phi_i}^\dagger \phi_i) \,,\,\,\,\,
c_i=  \mbox{Tr}({\phi_i}^\dagger \phi_i \phi_{i+1} {\phi_{i+1}}^\dagger) \,\,\, ,
\end{equation}
we study the potential
\begin{equation}
V_i = \tilde{m}^2 a_i + \tilde{\lambda}\, a_i^2 + \tilde{\lambda}' b_i + \tilde{\rho}\,
a_i\,a_{i+1}+\tilde{\eta} \, c_i \,\,\, ,
\label{eq:gensun}
\end{equation}
\begin{equation}
V_b = -\tilde{m}_1^2 \, a_1 + \tilde{\lambda}_1 \, a_1^2 \,\,\, ,
\label{eq:gensunb}
\end{equation}
where, again, boundary corrections have been added to trigger spontaneous symmetry breaking.
To facilitate our numerical analysis of the potential, we choose $N=3$, since SU(3) is the
smallest SU(N) group that retains many of same group theoretical properties of larger SU(N)
(for example, non-vanishing $f^{abc}$ and $d^{abc}$ constants).  It is possible
to duplicate the warp factors shown in Fig.~\ref{fig:3.1} in the non-Abelian case, provided
that we make the parameter identifications
\begin{equation}
\tilde{m}_1^2=m_1^2/3 \,\,\,\,\,
\tilde{\lambda}_1 = \lambda_1/9 \,\,\,\,\,
\tilde{m}^2=m^2/3  \,\,\,\,\,
9\tilde{\lambda}+3\tilde{\lambda}'=\lambda \,\,\,\,\,
9\tilde{\rho}+3\tilde{\eta}=\rho \,\,\,.
\end{equation}
For $\tilde{\lambda}=\tilde{\lambda}'=\lambda/12$ and $\tilde{\rho}=\tilde{\eta}=\rho/12$
we have found numerically that our warped extrema remain stable minima of
the enlarged potential Eqs.~(\ref{eq:gensun}) and (\ref{eq:gensunb}).   The results
of Fig.~\ref{fig:3.1} can thus be applied to study the gauge boson spectrum in both
Abelian and non-Abelian examples.

Gauge boson masses originate from the link kinetic terms
\begin{equation}
{\cal L}_m = \sum_j^{n-1} \, \mbox{Tr} \left(D_\mu \phi_j\right)^\dagger
\left(D^\mu \phi_j\right) \,\,\, ,
\end{equation}
which reduce to
\begin{equation}
g^2\sum_j^{n-1} v_j^2 \, \mbox{Tr} \left[ A_{j+1} \cdot A_{j+1}
-2\, A_{j+1} \cdot A_{j} + A_{j} \cdot A_{j} \right] \,\,\, ,
\end{equation}
where $g$ is the gauge coupling, or in matrix form
\begin{equation}
m^2_{gauge} = g^2 \left( \begin{array}{ccccccc}
\quad v^2_1 \quad &  \quad -v^2_1 \quad & \qquad & \qquad & \qquad & \qquad &\\
-v^2_1 & v^2_1+v^2_2  & -v^2_2  & & & & \\
& -v^2_2 & v^2_2+v^2_3  & -v^2_3 & & & \\
&      &          &     & \ddots & \\
&      &          &     &    & v^2_{n-2}+v^2_{n-1}  & - v^2_{n-1} \\
&      &          &     &    & -v^2_{n-1} & v^2_{n-1} \end{array} \right) \,\,.
\label{eq:gbmm}
\end{equation}
For the warped solutions shown in Fig.~\ref{fig:3.1}, it is straightforward to
evaluate the eigenvalues of Eq.~(\ref{eq:gbmm}) numerically.  Results for the Abelian,
$n=100$ model are shown in Fig.~\ref{fig:3.2}.

\EPSFIGURE{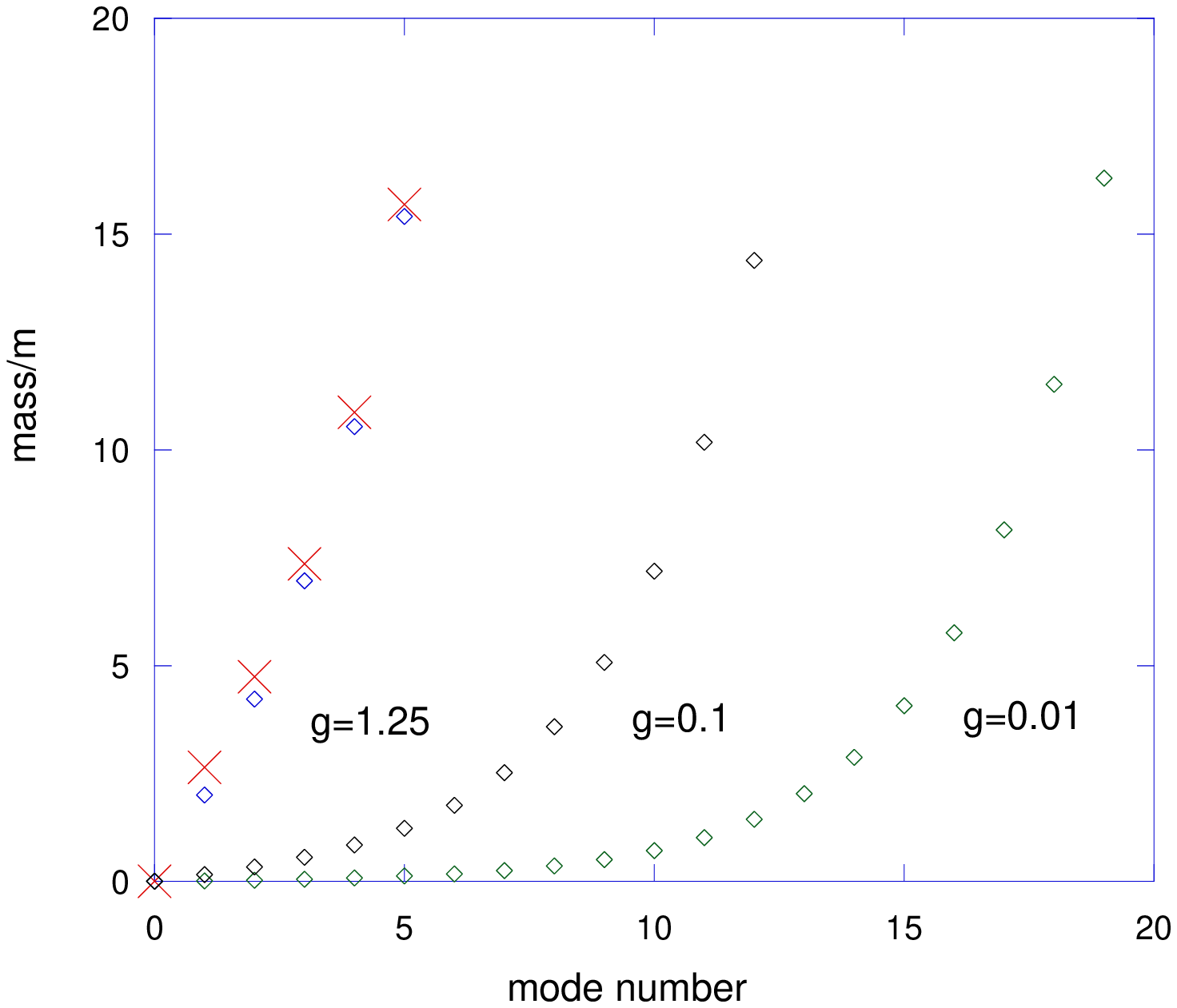,width=4.0in}{Mass spectra of the scalar and vector states for
the Abelian $n=100$ model, for $g=0.01$, $0.1$ and $1.25$.  Gauge (link) field states
are indicated by diamonds (crosses).\label{fig:3.2}}

A number of comments are in order.  While the gauge tower has a zero mode (corresponding to
the unbroken U(1) factor), the scalar tower has an ``almost" zero mode whose mass approaches
zero in the limit $n\rightarrow \infty$.  This mode can be identified as the pseudo-goldstone
boson of the broken approximate translation invariance of the moose.  In the Abelian models,
this state (as well as every other in its tower) is a gauge singlet and does
not necessarily portend any inescapable phenomenological problems.  However, more precise
statements can only be made in the context of specific phenomenological applications.
For the other scalar and vector modes, which we will label by an integer $k \geq 1$, the mass
spectra in Fig.~\ref{fig:3.2} are very accurately described by the exponential functions,
\begin{eqnarray}
m_k^{s} &=&  2.711\, m\, e^{0.347\,k} \nonumber \\
m_k^{v} &=&  2.086 g \, m\, e^{0.348\,k} \,\,\, ,
\end{eqnarray}
where $s$ and $v$ indicate the scalar and vector masses, respectively.  For the particular value
$g\approx 1.25$, the two towers of states become nearly degenerate.
In Reference \cite{Yael}, it was shown that the product of nonzero
eigenvalues of the mass matrix Eq.~(\ref{eq:gbmm}) is given by, \begin{equation}
\prod_i m_i^2 = N\prod_i g^2 v_i^2. \label{eq:Yael}\end{equation}
If the vevs $v_i$ vary exponentially, then (\ref{eq:Yael}) leads one to conjecture that
the masses may have an exponential spectrum for most of the tower, as we have found numerically.
Assuming a tower of the form $m_j\sim [\exp(-kR)/a] \exp(k a j)$, where $k$ is the
AdS scale and $a$ is the lattice spacing,
we expect the exponential to approximate the roughly linear tower
of the continuum theory for the first ${\cal O}(ka)$ modes.
It is useful to compare these results explicitly
to the spectrum of bulk scalar and vector modes in a 5D slice of
anti-de Sitter space.  Defining the parameter $x_n$ by
\begin{equation}
m_n = x_n \, k \,\exp(-k r_c\pi) \,\,\, ,
\end{equation}
where $k$ is the AdS curvature and $r_c$ is the compactification radius, the values of $x_n$ for
a massless bulk scalar are given by~\cite{gw}
\begin{equation}
2\, J_\nu (x_n) + x_n\, J'_\nu(x_n)=0 \,\,\,,
\label{eq:adsscalar}
\end{equation}
with $\nu=2$, and for a bulk U(1) gauge field by~\cite{dhr}
\begin{equation}
J_1(x_n)+x_n\, J'_1(x_n)+\alpha_n [ Y_1(x_n)+x_n \,Y'_1(x_n) ]=0 \,\,\, ,
\label{eq:adsvector}
\end{equation}
where $J$ and $Y$ are Bessel functions, and
\begin{equation}
\alpha_n \approx -\frac{\pi}{2} [ \ln (x_n/2)-kr_c \pi +\gamma + 1/2]^{-1} \,\,\, .
\label{eq:alphan}
\end{equation}
Note that $\gamma\approx 0.577$ is the Euler constant, and the Eqs.~(\ref{eq:adsscalar}),
(\ref{eq:adsvector}) and (\ref{eq:alphan}) are accurate provided that $\exp(-k r_c\pi) \ll 1$.
The values of $x_n$ can be obtained numerically, and increase approximately linearly with $n$.
For simplicity, we can compare this spectrum to our original, fine-tuned model where
$a_i \approx m^2 \lambda^{j-1}$, with $\lambda$ set to $1/2$. Assuming the phenomenologically
motivated value $k r_c \approx 11.27$
(to generate the Planck scale/weak scale hierarchy)
and the choice $m \approx 5\times 10^{-12} k$ (to match the light KK spectrum
of the continuum theory), we obtain
the first few modes shown in Fig.~\ref{fig:3.3}.

\EPSFIGURE{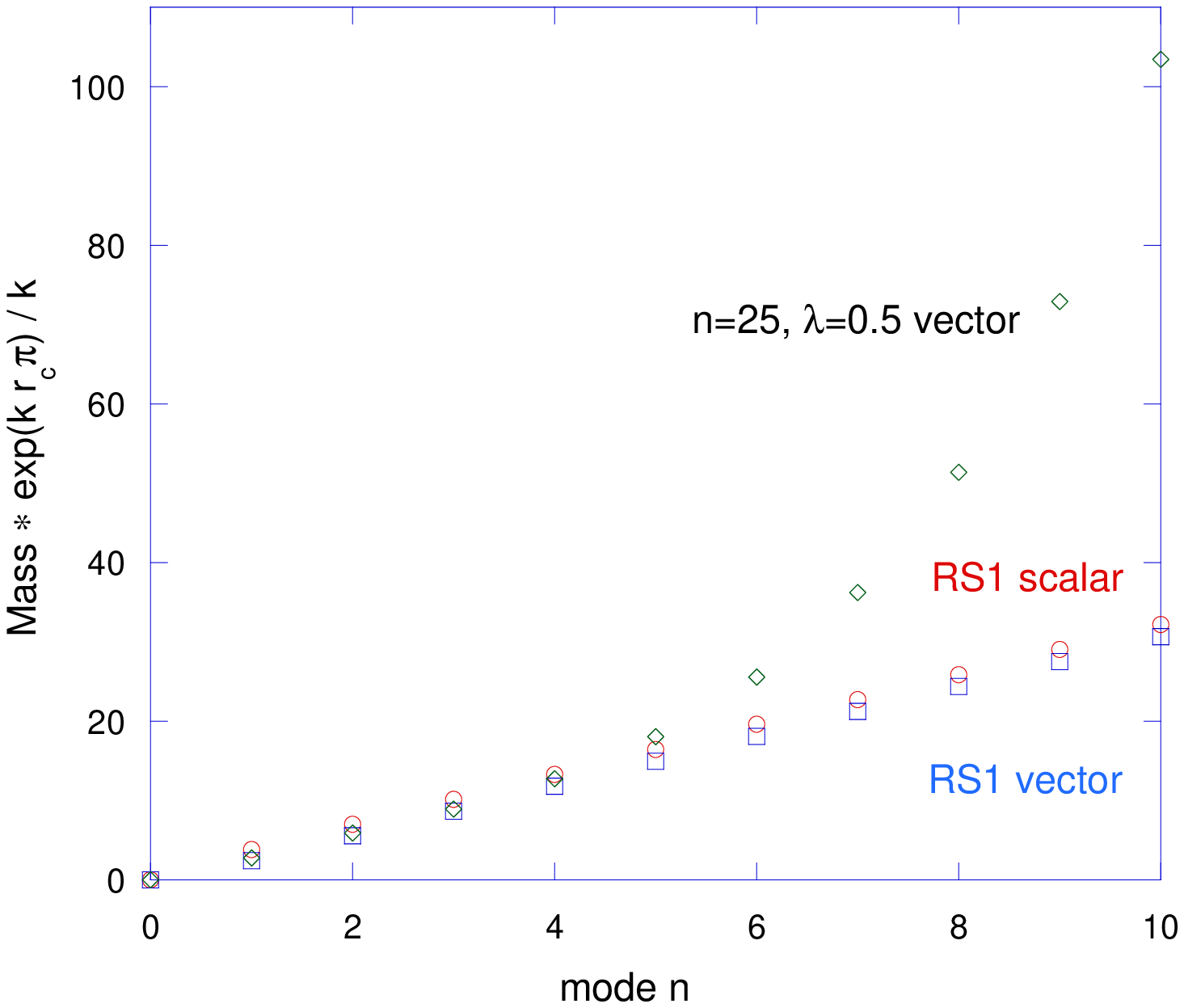,width=4.0in}{Comparison of the spectra of our simplified
deconstructed model to the continuum results Eqs.~(\ref{eq:adsscalar}) and
(\ref{eq:adsvector}), assuming $\lambda=1/2$, $m \approx 5\times 10^{-12} k$ and
$k r_c \approx 11.27$.  See the text for discussion.
\label{fig:3.3}}

It is therefore possible that the deconstructed models presented here can effectively mimic the
first few gauge Kaluza-Klein modes of the Randall-Sundrum model, even with a
coarse-grained lattice ({\em i.e.} large lattice spacing). As one would expect,
fine lattices do a better job of reproducing continuum results.  As an example,
let us assume the hierarchy $v_{n-1}/v_1=e^{-k\pi r_c}=e^{-30}$. Since $\pi r_c=na$
we identify the AdS scale, $k=(k\pi r_c)/(na)=30/(na)$. We also identify
the lattice spacing, $a^{-1}=g v_1$, where we have chosen to shift the warp factor so that
$f(y_1)=0$.  This choice corresponds to $\lambda=e^{-60/(n-1)}=0.985$ for
$n=4000$ in the parametrization of the mass matrix used earlier.
To summarize, the mapping of continuum parameters to lattice parameters in this model
is $k=30 gv_1/n$ and $\pi r_c=n/(gv_1)$. In Fig.~\ref{fig:finelattice} we consider
a relatively fine lattice with 4000 lattice sites and
compute the spectrum of gauge boson masses; there is relatively good agreement between
this result and that of the continuum.  Whether or not there is a continuum interpretation
of the models we consider over a large range of energies, our warped deconstructed models
are interesting in their own right, as we will describe in the following sections.

\EPSFIGURE{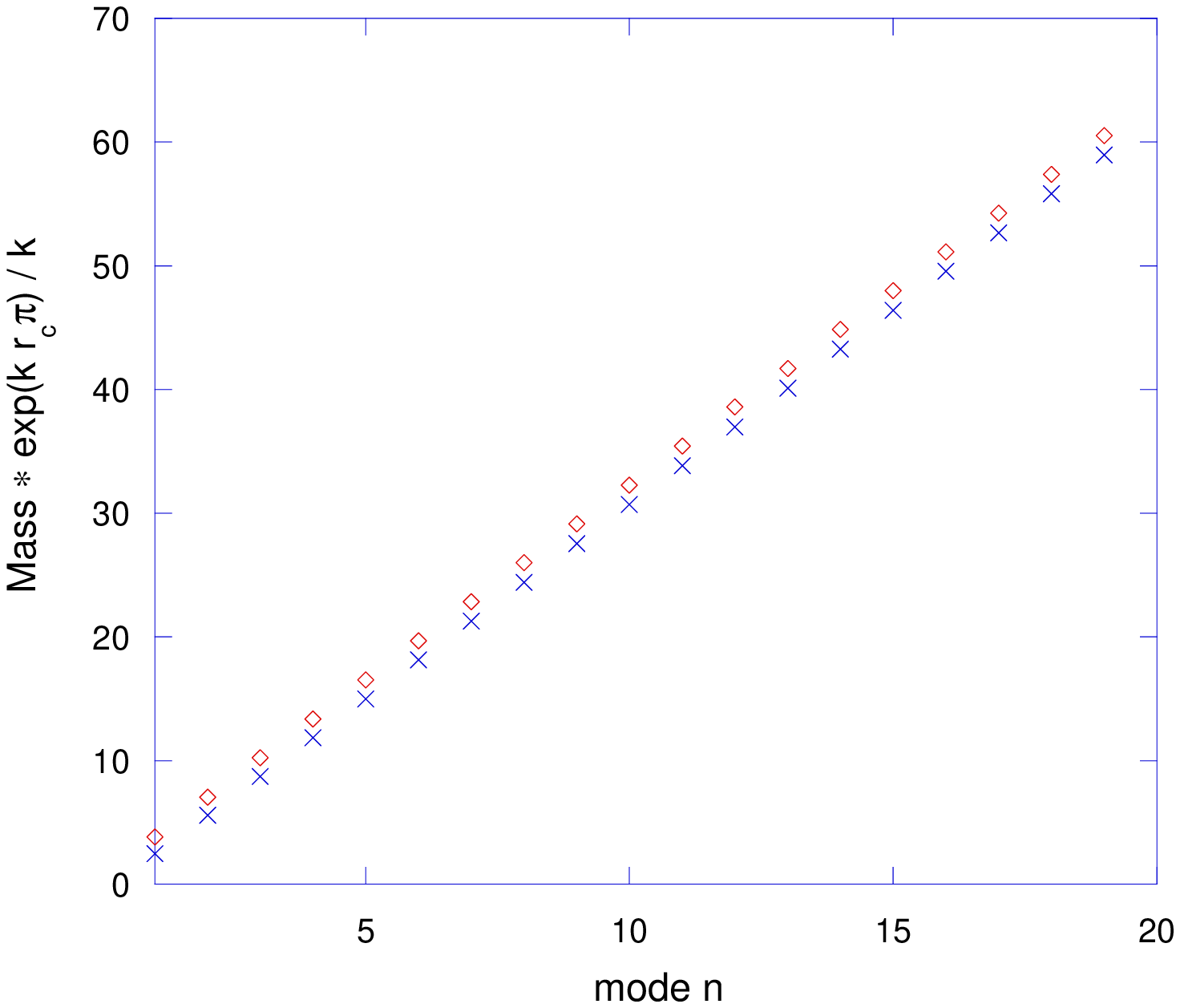,width=4.0in}{Comparison of the continuum (crosses) and the
deconstructed gauge boson spectrum for $4000$ lattice sites (diamonds), in
units of $k\exp(-k\pi r_c)$ with $k\pi r_c=30$.
\label{fig:finelattice}}

Finally, we comment on the scalar mass spectrum of the non-Abelian models. We have already
noted that the potential given in Eqs.~(\ref{eq:gensun}) and (\ref{eq:gensunb}) generate warp
factors, and we identified solutions that are minima of the potential.
We note here that the bifundamentals decompose under the diagonal gauge
group into a real adjoint which is eaten; and a real adjoint and complex
singlet which remain in the physical spectrum.
The uneaten fields do not necessarily have an
extra dimensional interpretation.  (The adjoints are necessary
in the supersymmetric version of the theory to form KK modes of a 5D SUSY
multiplet.)
The $n-1$ singlets under the diagonal
gauge group remain massless and are identified with the Goldstone bosons
of the spontaneously broken U(1)$^{n-1}$ global symmetry acting on the
$\phi_i$.  One may include SU(N)$_i$-invariant operators like
\begin{equation}
\Delta V_i = c \, M^{4-N} \, \det \phi_i + \mbox{h.c.}
\end{equation}
with small coefficient $c$, and the $n-1$ singlet states in question will become massive
without spoiling the pattern of link field vevs obtained in the absence of these terms.
It is also worth noting that higher-dimension operators that break the translation
invariance of the moose can be included to raise the mass of the lightest adjoint scalar
state.

\subsection{Gauge Coupling Running}
There are many possible applications of the
nonsupersymmetric warped deconstructed theories that we have
just considered.  For example, one could construct purely
four-dimensional analogs of the warped theories that attempt
to explain fermion masses via bulk wave function
overlaps~\cite{huber}.  Whatever the application, one is
bound to ask how the mass spectra described in the previous
section affect gauge coupling running and unification~\cite
{deconrun}.  We consider that issue in this section.

We begin with the generic observation that the towers of
gauge and link fields that we obtained were well
approximated by
\begin{equation}
m_0=0 \,\,\,\,\, \mbox{and}\,\,\,\,\, m_j = \bar{m} e^
{\bar{a}\,j},\, j \geq 1 \,
\end{equation}
when the number of sites was large. Here, the parameter set
$(\bar{m},\bar{a})$ corresponds to a particular towers of states,
and may differ between the gauge and uneaten link fields.
We will simplify our discussion by assuming that these
parameters are universal. However, as we noted earlier, the link
degrees of freedom that are not eaten by the gauge fields
could have a different spectrum.  While the gauge tower has
an exact zero mode, we assume that the lightest link field
modes (which are real scalar fields in the adjoint representations of the
diagonal, non-Abelian gauge groups) are heavy enough to evade direct detection, but
can be taken as massless as far as the renormalization group
running is concerned. This is equivalent to
saying that we ignore any low-scale threshold effects.

For simplicity, let us consider the effect of a single field with an exponential tower
of modes on the running of a diagonal gauge coupling. Imagine that we start at some
initial renormalization scale $\mu_0$ and evolve the gauge coupling $\alpha$ through
each KK mass threshold up to a scale $\mu$ that is given by
$m_N \leq \mu \leq m_{N+1}$.  One finds that
\begin{equation}
\alpha^{-1}(\mu)  =\alpha^{-1}(\mu_0) - \frac{b}{2\pi}
\ln\frac{m_1}{\mu_0}
-\sum_{j=2}^N \frac{j\, b}{2\pi}\ln\frac{m_j}{m_{j-1}} -
\frac{(N+1)\,b}{2\pi}
\ln\frac{\mu}{m_N} \,,
\label{eq:rge1}
\end{equation}
where $b$ is the one-loop beta function. The exponential
form of the spectrum for the massive modes in the KK towers
leads to a simplification of the logarithms in
Eq.~(\ref{eq:rge1}), which in turn allows us to do the sum
in the third term.  The result is
\begin{equation}
\alpha^{-1}(\mu)  =\alpha^{-1}(\mu_0) - \frac{b}{2\pi}
\ln\frac{\mu}{\mu_0}
-\frac{N\,b}{2\pi}\ln\frac{\mu}{\bar{m}} + \frac{\bar{a}\,b}{4\pi}
N(N+1) \,\,\,.
\label{eq:rge2}
\end{equation}
The first two terms give the usual one-loop renormalization
group running of the couplings between the scales $\mu_0$
and $\mu$; the last two terms are corrections due to
the particular form of our KK towers.  To understand the
effect of these terms, it is useful to note that for large
$N$, $N(N+1) \approx N^2 \approx (\ln\frac{\mu}{\bar{m}}/\bar{a})^2
$.  Thus, unlike gauge coupling running in the standard
model, Eq.~(\ref{eq:rge2}) has a quadratic
dependence on the log of the renormalization scale.

This point has been noted before in studies of gauge
coupling running in deconstructed AdS
space~\cite{deconrun}.  The presence of log squared terms
arises due to the choice of boundary conditions on the gauge
couplings.  In our models, we define the gauge couplings to
have a common value at a common scale, which can be
identified as the scale of the highest link
field vev.  This choice is required by the assumed
translation invariance of our theories.
However, to reproduce the purely logarithmic gauge coupling
evolution expected in AdS space,
one must define each gauge coupling of the deconstructed
theory at the scale of the corresponding link vev, before
setting the couplings equal~\cite{deconrun}.  In the
framework that we have presented, there is no symmetry of
the four-dimensional theory that would make
such a choice natural.  We therefore use Eq.~(\ref{eq:rge2})
to draw our phenomenological conclusions.

Let us now apply Eq.~(\ref{eq:rge2}) to the standard model.
We take $\mu_0=m_Z$ and the gauge couplings $\alpha^{-1}
_1=59.02$, $\alpha^{-1}_2=29.57$ and $\alpha^{-1}_3=8.33$.
Our SU(3), SU(2) and U(1) beta function contributions are
\begin{equation}
\begin{array}{lcl}
\mbox{gauge (massless)}  &\mbox{   } & (-11,-22/3,0) \\
\mbox{gauge (massive)} & &  (-21/2,-7,0) \\
\mbox{physical links}  & &(1/2, 1/3, 0) \\
\mbox{matter} & &(4 , 4, 4) \\
\mbox{Higgs}  & &(0, 1/6, 1/10)
\end{array}\label{eq:bfnc}
\end{equation}
One sees that the sum of massless gauge, higgs and matter beta functions in
Eq.~(\ref{eq:bfnc}) is $(-7, -19/6, 41/10)$, the usual result in the standard model
with one electroweak Higgs doublet. As a further check, one can also note that the sum of
Higgs plus massive gauge beta functions is $(-21/2,-41/6,1/10)$, which agrees with the
KK beta function given in Ref.~\cite{power-law} for the nonsupersymmetric standard model
with only the gauge fields and one Higgs doublet in the bulk.  In the present application,
the beta functions multiplying the $\ln(\mu/\mu_0)$ term in Eq.~(\ref{eq:rge2}) are the
sum of the Higgs, matter, physical link, and massless gauge beta functions shown in
Eq.~(\ref{eq:bfnc}), $(-13/2,-17/6, 41/10)$; the beta function contributions of each KK level
is the sum of the physical link and massive gauge beta functions, $(-10,-20/3,0)$. As an
example, the choice $\bar{a}=1$ and $\bar{m}=1$~TeV leads to unification at the scale
$7\times 10^5$~GeV with
\begin{equation}
\frac{\alpha^{-1}_3 - \alpha^{-1}_{12}}{\alpha^{-1}_{12}} =
11.8 \%,
\end{equation}
where $\alpha^{-1}_{12}$ is evaluated at the point where the
SU(2) and U(1) couplings unify.  This is not terribly
impressive, but should be put in the appropriate context.
Unification in the nonsupersymmetric standard model occurs
at $1 \times 10^{13}$~GeV with
\begin{equation}
\frac{\alpha^{-1}_3 - \alpha^{-1}_{12}}{\alpha^{-1}_{12}} =
13.4 \%
\end{equation}
using the input numbers defined earlier.  Thus, the
existence of the exponential towers of gauge and link states
actually improves unification slightly in comparison
to the minimal standard model.  This is significant since we
had no reason {\em a priori} to expect that approximate
unification would be possible at all. From a practical point
of view, this suggests that any of a number of possible
corrections to standard model unification (for
example, those motivated by split supersymmetry) could
correct this result as needed.  We do
not pursue this possibility further here.

\section{Deconstructed Warped SUSY and Collective SUSY Breaking}
\label{sec:susy}

In this section we consider models with global supersymmetry.
We dynamically deconstruct a warped 5D supersymmetric U(1)
gauge theory, and discuss the unusual properties and phenomenology
of this model.  Related models were studied in \cite{pokorski,dienes,japanese}.

\subsection{The Deconstructed SUSY U(1) Theory}
The 4D theory is an ${\cal N}$=1 supersymmetric moose theory. Unlike our previous
examples, the U(1) gauge fields now belong to vector multiplets with the associated gauginos,
and the link fields $\phi_i$ are chiral multiplets with charges $(+1,-1)$
under neighboring gauge groups. To avoid gauge anomalies we can use the
Green-Schwarz mechanism or add Wess-Zumino terms.  The structure of the anomaly-canceling
sector of the theory is tightly constrained if the theory is to appear extra
dimensional in the limit of small lattice spacing \cite{pokorski}.
We will not concern ourselves with the details of this sector of the theory,
but rather cancel anomalies by introducing a duplicate set of link fields
$\overline{\phi}$, but with opposite charges $(-1,+1)$, resulting in the quivers of
Figure~\ref{fig:susyquiver}.  The additional bifundamentals have no extra
dimensional interpretation, but they also are singlets under
the low-energy U(1) gauge symmetry and for most practical purposes can be ignored.
In the following, we will include the doubled set of link fields with the understanding
that one of the two sets will not have a higher-dimensional interpretation.
\EPSFIGURE{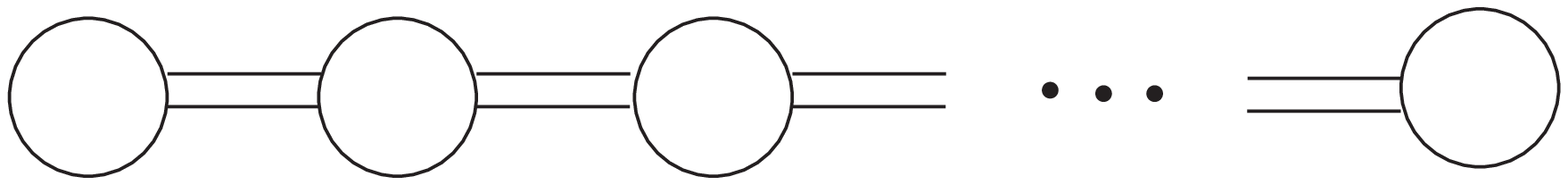,width=4in}{Moose for the supersymmetric U(1)$^n$ theory.
Doubled links indicate chiral superfields with charges $(+1,-1)$ and $(-1,+1)$.
 \label{fig:susyquiver}}

Warping of the extra dimensions in the Abelian theory can
be accomplished through the addition of Fayet-Iliopoulos (FI) terms $\xi_i$.
The potential for the scalar link fields arises from the D-terms, and is given
by: \begin{equation}
V_D=\sum_{i=1}^n D_i^2,
\end{equation}
where, \begin{equation}
D_i = g\left( |\phi_{i}|^2 - |\phi_{i-1}|^2 -
|\overline{\phi}_{i}|^2 + |\overline{\phi}_{i-1}|^2 + \xi_{i}\right).
\label{eq:D-terms}
\end{equation}
As usual, for a circular moose we define
$\phi_0=\phi_n$ and for the interval (orbifold) moose we define $\phi_0=\phi_{
n}=0$ (and similarly for $\overline{\phi}_0$ and $\overline{\phi}_n$).
We will again focus on the orbifold theory. We assume for now that the superpotential
vanishes, so that the only contribution to the scalar potential is due to the D-terms.

The stationary points of the D-term potential satisfy, \begin{equation}
\left<\phi_i\right>\,\left(\left<D_i\right>-\left<D_{i+1}\right>\right)=0\ .
\end{equation}
The vacua generically have equal D-terms, with, \begin{equation}
\left<D_i\right>=\frac{\sum_j g\xi_j}{n}\equiv D. \label{eq:Dvac}\end{equation}
As a result, the scalar VEVs $v_i$ and $\overline{v}_i$
satisfy the recursion relation, \begin{equation}
(|v_{i+2}|^2-|\overline{v}_{i+2}|^2)-
2(|v_{i+1}|^2-|\overline{v}_{i+1}|^2) +
(|v_i|^2 - |\overline{v}_i|^2) = (\xi_{i+1}-\xi_{i+2})\ .
 \label{eq:D-eqn}\end{equation}
The U(1)$^n$ gauge
symmetry is generically broken to a diagonal U(1).  The second
set of link fields does not alter this symmetry breaking pattern because the
link fields are neutral under the unbroken U(1).  There are flat directions
in the potential for which $|\phi_i|^2$ and $|\overline{\phi}_i|^2$ are shifted
by the same constant $c_i$.  These flat directions correspond to the
$n-1$ moduli $\overline{\phi}_i \phi_i$.
For simplicity in what follows, we will assume
$\overline{v}_i=0$, and we will use the gauge symmetry to make the $v_i$ real.
None of the following results changes qualitatively if we allow for
$\overline{\phi}_i$ vevs. Alternatively, as discussed earlier, we
can remove the $\overline{\phi}$ chiral multiplets from the theory and include
Wess-Zumino terms in the action to cancel the gauge anomalies.

One amusing consequence of supersymmetry in this theory is that the spectrum of massive
chiral multiplets is the same as the spectrum of massive vector multiplets, as required
in order to mimic the KK spectrum of a supersymmetric 5D gauge theory.   The SUSY Higgs
mechanism  forces the scalar masses to equal the gauge boson masses, resulting in a 4D
${\cal N}$=2 supersymmetric spectrum of massive fields \cite{CEGK}.

Notice that the relation between scalar VEVs (\ref{eq:D-eqn}) is a
latticized form of the equation, \begin{equation}
\frac{\partial^2|\phi(y)|^2}{\partial y^2} = -\frac{\xi'(y)}{a},
\label{eq:Dvac2}\end{equation}
where $a$ is a lattice spacing that will be defined in terms of the
fundamental parameters of the theory shortly.  The explicit ultraviolet
dependence in the continuum scalar field equation could be absorbed in a
redefinition of the FI terms, but we will not do that here.
Equation~(\ref{eq:Dvac2}) can be integrated once to give,
\begin{equation}
\frac{\partial|\phi(y)|^2}{\partial y} = \frac{-\xi(y)+D/g}{a},
\label{eq:firstderiv}\end{equation}
with integration constant $D/g$, which is the continuum form of (\ref{eq:Dvac})
with $D_i$ given by (\ref{eq:D-terms}) and
\begin{equation}
D/g=\int_0^R dy \,\xi(y)/R. \end{equation}
Equation (\ref{eq:firstderiv}) relates
the warp factor of (\ref{eq:metric}) to the 4D Fayet-Iliopoulos terms:
\begin{equation}
\frac{\partial e^{-f(y)}}{\partial y}=(-g^2 \xi(y)+gD)a. \end{equation}

The warp factors that can be obtained in this way form a restricted class.
As a particular example, if all of the FI terms are equal then from
(\ref{eq:firstderiv}) with $\xi(y)=D/g=const$,
the warp factor is constant and the metric reproduces flat spacetime.
As another example, if the first of the FI terms differs from the rest, then the right-hand
side of (\ref{eq:firstderiv}) is constant in $y$ except at the special site with
unique FI term (corresponding to a delta function in the continuum limit).
The resulting (squared) warp factor $e^{-f(y)}$ has a linear profile.

More generally, we note that
according to (\ref{eq:Dvac}) each D-term
is equal to the average value of $g\xi_j$, which in the continuum limit becomes
\begin{equation}
D=g\int_0^R dy\,\frac{\xi(y)}{R}.
\end{equation}
Then, suppose we want to fix the right-hand side of (\ref{eq:firstderiv})
so as to reproduce a particular warp factor, so that
\begin{equation}
\xi(y)=\widetilde{\xi}(y)+D/g = \widetilde{\xi}(y) +\int_0^R dy\, \xi(y)/R,
\label{eq:inteq}\end{equation}
for some specified profile $\widetilde{\xi}(y)$.  Whether or not there is a solution to the
integral equation (\ref{eq:inteq}) depends on the choice of $\widetilde{\xi}(y)$.  To determine
the constraint on $\widetilde{\xi}(y)$ we integrate (\ref{eq:inteq}) over $y$ and find,
\begin{equation}
\int_0^R dy\,\widetilde{\xi}(y)=0.
\end{equation}
Hence, we learn that in order to obtain a monotonic warp factor, there must be
a delta function contribution to $\xi(y)$ at a boundary of the spacetime.
We can also see the difficulty in obtaining a monotonic warp factor
by recognizing that $\widetilde{\xi}(y)$ is the difference
between $\xi(y)$ and its average value over the interval $(0,R)$.
As a result, if $\widetilde{\xi}(y)>0$ for some y,
then there must exist some $y'$ where $\xi(y')<0$.  The same argument applies to the latticized
theory: If for some $i$, $(\xi_i-D/g)>0$ then there exists an $i'$ for which
$(\xi_{i'}-D/g)<0$.  Then, by Eqs.~(\ref{eq:D-terms}) and (\ref{eq:Dvac}), in order for the
warp factor to be monotonic, the FI term $\xi_1$ at the boundary will differ in sign from
the FI terms in the bulk.  (In fact, we will see
in Section~\ref{sec:SUSYbreaking} that $\widetilde{\xi}(y)$ is the
profile of FI terms in an equivalent theory with vanishing vacuum energy.)

\subsection{The KK spectrum}
The masses of the components
of the vector and chiral multiplets arise from the K\"{a}hler potential for
the bifundamental chiral multiplets, \begin{equation}
{\cal L}\supset \int d^4\theta\,
\sum_i \Phi_i^\dagger\,\exp\left[g(V_{i}-V_{i+1})\right]\,\Phi_i +
\overline{\Phi}_i^\dagger\, \exp\left[g(-V_{i}+V_{i+1})\right]\,\overline{\Phi}_i,
\end{equation}
where $\Phi_i$ is the bifundamental chiral multiplet charged under
U(1)$_{i+1}$ and U(1)$_i$, and $V_i$ is the U(1)$_i$ vector multiplet.
We will separately calculate
the gauge boson, scalar, and fermion masses, and find that the spectrum is
supersymmetric despite the nonvanishing D-terms in the vacuum.  Later we
will explain why the presence of global supersymmetry  is to be expected,
and we will study the unusual SUSY breaking phenomenology of this and
related models.

\subsubsection{Gauge bosons}

On the supersymmetric orbifold, as in the nonsupersymmetric case,
the gauge boson masses arise from the bifundamental vevs through the
Higgs mechanism, with mass terms, \begin{equation}
{\cal L}\supset
\frac{1}{2} A_{i \,\mu}M_{ij}A_j^\mu \ .\end{equation}
The mass-squared matrix is, as before,\begin{equation}
m_{\rm gauge}^2 = 2 g^2 \left(\begin{array}{ccccccc}
v_1^2 && -v_1^2 && &&  \\
-v_1^2 && v_1^2+v_2^2 && -v_2^2 && \\
 && -v_2^2 && v_2^2+v_3^2 &-v_3^2 & \\
 &&&& \ddots & \ddots & \\
 && &&&    -v_{n-1}^2 &\  v_{n-1}^2 \end{array} \right)
\label{eq:sgbmm}\end{equation}

Identifying the lattice spacing with $a\equiv 1/(gv_1)$ as in the
nonsupersymmetric theory, we recover the spectrum of gauge fields in a
latticized warped extra dimension with metric, \begin{equation}
ds^2=e^{-f(y)} dx^2+dy^2 \ ,\end{equation}
with warp factor $e^{-f(na)}=v_n/v_1$.  The action of the continuum theory
then includes, \begin{equation}
S\supset \int d^5 x \left(-\frac{1}{4}F_{MN}F^{MN}\right). \end{equation}
The relative factor of two between
Equations (\ref{eq:sgbmm}) and (\ref{eq:gbmm}) is due to our normalization of
the generators in the Abelian and non-Abelian theories.  We have chosen
$\rm{Tr}\,T^a T^b = c \delta_{ab}$ with $c=1/2$ for the non-Abelian theory, but
$c=1$ for the Abelian theory.

\subsubsection{Fermions}
The K\"ahler potential for the chiral multiplets couples
the  chiral multiplets to the gauginos, and gives rise to the following mass
terms in the fermion Lagrangian:
\begin{eqnarray}
{\cal L}&\supset& i\frac{g}{\sqrt{2}}\,\sum_i \left[
\lambda_i(v_i \psi_i-v_{i-1}\psi_{i-1})
-\overline{\lambda}_i(\overline{\psi}_i-\overline{\psi}_{i-1})\right]\ \\
&=&
i\frac{g}{\sqrt{2}}\, (\lambda_i\ |\ \psi_i)\ \left(\begin{array}{c|c} & \ \Theta \\ \hline
\Theta^\dagger \ & \end{array}\right)_{ij}\ \left(\begin{array}{c}\lambda_j \\ \hline \psi_j
\end{array}\right)\ +\ {\rm h.c.}, \end{eqnarray}
where the $n\times(n-1)$ dimensional matrix $\Theta$ is, \begin{equation}
\Theta=\left(\begin{array}{cccc}
v_1 & & & \\
-v_1 & v_2 & & \\
 & \ddots & \ddots & \\
 & & -v_{n-2} & v_{n-1} \\
 & & & -v_{n-1} \end{array} \right)\ . \end{equation}

The fermion mass matrix is identified by writing the fermion Lagrangian in
the form, \begin{equation}
{\cal L}\supset
\frac{1}{2}\, (\lambda_i\ |\ \psi_i)\ M_{ij}
\left(\begin{array}{c}\lambda_j \\ \hline \psi_j
\end{array}\right)\ +\ {\rm h.c.} \end{equation}
The squared mass matrix is the given by, \begin{equation}
M^2_{\rm fermions}=2 g^2
\left(\begin{array}{cc} \Theta \Theta^\dagger &  \\
&  \Theta^\dagger \Theta
\end{array} \right)\ .\end{equation}
The block diagonal elements of the mass matrix are proportional to,
\begin{eqnarray}
\Theta\Theta^\dagger &=& \left(\begin{array}{ccccc}v_1^2 & -v_1^2 &&& \\
-v_1^2 & \ v_1^2+v_2^2 & -v_2^2 && \\
 & -v_2^2 &  v_2^2 +v_3^2 & -v_3^2 & \\
 & & \ddots & \ddots &   \\
 & & & -v_{n-1}^2 & v_{n-1}^2 \end{array}\right) \,, \nonumber \\
\Theta^\dagger \Theta&=&
\left(\begin{array}{ccccc}2v_1^2 & -v_1 v_2 &&& \\
-v_1 v_2 & 2v_2^2 & -v_2 v_3 && \\
 & -v_2 v_3 & 2v_3^2 & -v_3 v_4 & \\
 & & \ddots & \ddots &  \\
 & & & -v_{n-2} v_{n-1} & 2v_{n-1}^2 \end{array}\right)\ .
\label{eq:fermimass}\end{eqnarray}
The upper-left diagonal block of the mass squared matrix $M^2_{\rm fermions}$ is
identical to the gauge boson mass matrix.  The bottom right
diagonal block has identical eignenvalues
to the first, except that the zero mode is missing from that sector.  The
single fermion zero mode is therefore composed entirely of the gauginos:
 \begin{equation}
\lambda\equiv \frac{1}{\sqrt{n}} \sum_{i=1}^n \lambda_i.
\label{eq:Goldstino}\end{equation}
This zero mode  will be important later in the discussion of SUSY
breaking.  The massive modes match the spectrum of massive gauge bosons, as
required for 5D supersymmetry.

\subsubsection{Scalars}
The scalar masses are determined by the D-term potential
Eq.~(\ref{eq:D-terms}).  Defining $d_i\equiv \left<D_i\right>=
v_i^2-v_{i-1}+\xi_i$, and $\phi_i=v_i+\varphi_i$, we may expand,
\begin{eqnarray}
V_D&=& g^2\sum_i\left[(\varphi_{i+1}+v_{i+1})(\varphi_{i+1}^\dagger+v_{i+1})-
(\varphi_{i}+v_{i})(\varphi_{i}^\dagger+v_{i})-|\overline{\phi}_{i+1}|^2+
|\overline{\phi}_{i}|^2+\xi_{i+1}\right]^2 \nonumber\\
&=&g^2\sum_i\left[d_{i+1}+|\varphi_{i+1}|^2+v_{i+1}\,(\varphi_{i+1}^\dagger+
\varphi_{i+1})-|\varphi_{i}|^2-v_{i}\,(\varphi_{i}^\dagger+
\varphi_{i})-|\overline{\phi}_{i+1}|^2+
|\overline{\phi}_{i}|^2\right]^2 \,,\nonumber
\end{eqnarray}
from which if follows that
\begin{eqnarray}
V_D \supset 2g^2\sum_i (d_{i}-d_{i+1})&&\left(|\varphi_{i}|^2+v_{i}\,
(\varphi_{i}^\dagger+\varphi_{i})-|\overline{\phi}_{i}|^2\right) \nonumber \\
&&+g^2\sum_i \left(v_{i+1}(\varphi_{i+1}^\dagger+\varphi_{i+1})-v_i
(\varphi_i^\dagger+\varphi_i)\right)^2 \nonumber \\
=2g^2\sum_i (d_{i}-d_{i+1})&&\left(|\varphi_{i}|^2+v_{i}\,
(\varphi_{i}^\dagger+\varphi_{i})-|\overline{\phi}_{i}|^2\right) \nonumber \\
\nonumber \\
&&+\frac{1}{2}(\varphi_i^\dagger \ |\ \varphi_i)\,g^2\left(\begin{array}{c|c}\Theta^\dagger
\Theta \ & \ \Theta^\dagger \Theta \nonumber \\ \hline
\Theta^\dagger \Theta\  & \ \Theta^\dagger \Theta \end{array}\right)
\left(\begin{array}{c}\varphi_i \nonumber
\\ \hline \varphi_i^\dagger \end{array}\right)\ , \label{eq:scalarmass}
\end{eqnarray}
where $\Theta^\dagger\Theta$ is the same matrix that determines the fermion masses
Eq.~(\ref{eq:fermimass}).  We will see again in Section~\ref{sec:SUSYbreaking} that
the D-terms in the vacuum, $d_i$, are all equal, so that the terms in (\ref{eq:scalarmass})
proportional to $(d_i-d_{i+1})$ vanish.

Diagonalizing the mass matrix, the imaginary modes
$(\varphi_i-\varphi_i^\dagger)$ have vanishing masses, and are the $n-1$
eaten Goldstone modes of the U(1)$^n\rightarrow$U(1) gauge symmetry breaking
pattern.  The real modes have the same masses as the massive fermions
and gauge bosons.  The $\overline{\phi}_i$ fields remain massless, and do not have
a higher-dimensional interpretation.

The 5D vector multiplet is decomposed into a tower of massive 4D vector multiplets,
consisting of a massive gauge boson, two Weyl fermions and a real scalar;  it
includes, in addition,  a massless vector multiplet and chiral multiplet, consisting
of a massless gauge boson, two Weyl fermions, and a complex scalar. In the continuum
theory, by allowing the orbifold action on the fields to include
a transformation by an element
of the R-symmetry of the supersymmetric theory, supersymmetry can
be partly or completely broken. The ``boundary conditions,'' specified by
the terms in the Lagrangian from the first and last site of the moose,
project out
one massless Weyl fermion and one massless complex scalar, leaving an
${\cal N}$=1 vector multiplet zero mode consisting of a massless gauge
boson and Weyl fermion.  (We could have kept an extra zero-mode chiral
multiplet by adding an additional chiral multiplet charged under the first
U(1) gauge group in the moose and not giving its scalar component a vev.)
Hence, we have found a spectrum of gauge bosons, fermions and scalars
consistent with 5D supersymmetry partially broken to 4D SUSY at the zero-mode level
by the orbifold boundary conditions.  The U(1) D-terms do not vanish in the
vacuum, so the fact that we have recovered a supersymmetric low-energy theory
may at first sight seem surprising.  This is the subject of the next section.

\subsection{SUSY Without Supergravity} \label{sec:SUSYbreaking}
Unless the  FI terms are fine-tuned such that $\sum_i \xi_i=0$, then by
Eq.~(\ref{eq:Dvac}) the D-terms cannot simultaneously vanish and it
would seem that there should then be a collective breaking of supersymmetry
from the nonvanishing D-term VEVs. In fact this is not the case, as we have seen
that the spectrum of the deconstructed U(1) orbifold
theory preserves four supercharges (${\cal N}$=1 in 4D).  The reason this
is possible is that only the unbroken diagonal U(1) has a nonvanishing FI term,
\begin{equation}\xi_{\rm diag}=\frac{1}{\sqrt{n}} \sum_i \xi_i\ . \end{equation}
All linear combinations of U(1)'s orthogonal to the diagonal U(1)
have vanishing FI terms by orthogonality and the equality of the
D-term VEVs $d_i$.

The diagonal U(1) has no charged matter, so the FI term
is just a cosmological constant and can be shifted away (until we couple
the theory to gravity).  To see this explicitly, we define, \begin{equation}
\widetilde{\xi}_i\equiv \xi_i-\frac{\sum_i \xi_i}{n}\ , \label{eq:xi-tilde}
\end{equation} and
\begin{equation}
\widetilde{D}_i \equiv g\left( |\phi_{i}|^2 - |\phi_{i-1}|^2 -
|\overline{\phi}_{i}|^2 + |\overline{\phi}_{i-1}|^2 + \widetilde{\xi}_{i}\right)\ ,
\label{eq:D-tilde}
\end{equation}
after which the D-term potential becomes \begin{equation}
V_D=\sum_i \widetilde{D}_i^2 +\frac{\left(\sum_i \xi_i\right)^2}{n}\ .
\end{equation}
We have defined $\widetilde{\xi}_i$ such that $\sum_i \widetilde{\xi}_i=0$.
Hence, the deconstructed U(1) theory with arbitrary FI terms is equivalent
to a theory with FI terms $\widetilde{\xi}_i$ in which global SUSY is unbroken
(according to Eq.~(\ref{eq:Dvac})) plus a cosmological constant $\Lambda=(\sum_i \xi_i)^2/n$.

According to the usual definition of the supersymmetry transformations,
the Goldstino transforms non-homogeneously.  This is usually taken to
be the indicator of supersymmetry breaking.  However, in this theory we
can redefine the action of the SUSY generators on the gauginos such that
the non-homogeneous part of the SUSY transformed Goldstino is shifted away.
The Goldstino is identified with the massless fermion mode of
(\ref{eq:Goldstino}), $\lambda=\sum_i \lambda_i/
\sqrt{n}$.  The Goldstino
is composed entirely of gauginos, as expected in the absence of
F-terms.  Its SUSY transformation is,
\begin{equation}
\delta \lambda = i \epsilon
\frac{\sum_i D_i}{\sqrt{n}}+\sigma^{\mu\nu}\epsilon
\frac{\sum_i{F_{\mu\nu}^i}}{\sqrt{n}}\ , \end{equation}
where $\epsilon$ is the superspace parameter and $\sigma^{\mu\nu}=1/4\,[
\sigma^\mu,\sigma^\nu]$ with $\sigma^\mu$ the Pauli matrices for $\mu=1,2,3$
and $\sigma^0=-1$.
The vacuum value of the D-term part of the transformation rule makes the
gaugino transform non-homogeneously, which is usually taken to be the
indicator of SUSY breaking.
However, if we shift the
D-terms by their VEVs, we find that the following SUSY transformation is also
preserved by this theory:
\begin{equation}
\widetilde{\delta} \lambda =
i\frac{\sum_i \widetilde{D}_i}{\sqrt{n}}+\sigma^{\mu\nu}\epsilon
\frac{\sum_i{F_{\mu\nu}^i}}{\sqrt{n}}
\ . \end{equation}
The vacuum part of the $\widetilde{\rm D}$-term part vanishes by
definition, so that the would-be Goldstino
transforms homogeneously under the shifted SUSY transformation.
The SUSY transformation rule for the shifted auxiliary fields $\widetilde{D}_i$
is the same as for the original auxiliary fields $D_i$, \begin{equation}
\widetilde{\delta} \widetilde{D}_i=\delta D_i = \overline{\epsilon}\,
\overline{\sigma}^\mu \partial_\mu\lambda_i-\epsilon\sigma^\mu\partial_\mu
\overline{\lambda}_i. \end{equation}
We are free to shift the supersymmetry transformations in this way because
the diagonal D-term is independent of the matter fields in this theory,
and its SUSY transformation
depends only on derivatives of the gauginos so that the additional
shift of the gauginos by a constant does not affect the transformation of
the D-term.
Once again,
we are only able to shift the SUSY transformations in this theory because
there is no charged matter under the diagonal U(1) gauge symmetry which
has nonvanishing D-term.  Otherwise the shifted SUSY would not be a
symmetry of the theory because of the dependence of the auxiliary D-term on the
charged fields.  Also, when we couple the theory to gravity we will
no longer be able to redefine the supersymmetry in a similar way
because the positive vacuum energy breaks local supersymmetry.

\subsection{SUSY-breaking Phenomenology}
Despite the unbroken global SUSY identified above, SUSY breaking reappears
when the supersymmetric U(1) theory is coupled to gravity or a messenger sector.
In this case, one finds the unusual feature that the scales of SUSY breaking in the
messenger and in the gravity sector may be hierarchically different.
This result is obtained because the supersymmetry is broken collectively: if any one of
the D-terms conditions were removed, then there would be a supersymmetric
ground state.  The supersymmetry breaking is therefore nonlocal in nature
and, from the 5D perspective, corresponds to including an explicit
position-dependent cosmological constant in the theory by hand.  However,
one should not take this interpretation too literally, as we have not
deconstructed 5D gravity.  It should be clear, though, that this situation
is unlike models of deconstructed Scherk-Schwarz SUSY breaking
or radion F-term SUSY breaking \cite{kaplan-weiner}.  In those cases the
boundary conditions break global supersymmetry, while here we preserve
supersymmetry without supergravity.

If we are willing to give up on the higher-dimensional interpretation of the model,
then  we may assign random Fayet-Iliopoulos terms to the $n$ U(1) factors in the
theory.  The central limit theorem leads to an expected value of
$\sum_i \xi_i$ which grows like $\sqrt{n}$ for large number of lattice
sites.  Now we can imagine coupling one of the U(1) factors in the moose
to the MSSM through messenger fields.  If we define the typical scale for the $\xi_i$
to be $\hat{\xi}$ and take $g\sqrt{\hat{\xi}}\sim M_{Pl}$ then the SUSY breaking scale
as seen by the messengers and the splitting of messenger masses would be of size,
\begin{equation}
M_{\rm SUSY}^2=\left<D_i\right>= \frac{g^2\sum_i\xi_i}{n}\sim \frac{g^2\hat{\xi}}{n^{1/2}}\sim
\frac{M_{Pl}^2}{n^{1/2}}\ . \label{eq:Msusy}
\end{equation}
Gravity, on the other hand,
 would communicate a SUSY breaking scale specified by the cosmological
constant, \begin{equation}
M_{\rm grav}^2=\left(\sum_i D_i^2\right)^{1/2}\sim
g^2 \xi \sim M^2_{Pl}\ ,\end{equation}
and gravitino mass, \begin{equation}
m_{3/2}=M_{\rm grav}^2/M_{Pl}\sim M_{Pl}\ . \end{equation}

The suppression of the collectively broken SUSY
scale by the number of lattice sites in (\ref{eq:Msusy}) is reminiscent of
the suppressed scalar mass in little Higgs
theories~\cite{little-higgs}, although in the present case the suppression is
a tree level effect.  It is also reminiscent of a related model,
studied in Ref.~\cite{pokorski}, in which SUSY is broken by a deconstructed
Wilson line and Fayet-Iliopoulos terms at the boundary.  The collective
SUSY breaking in that model gives rise to a suppressed
SUSY breaking scale at the first lattice site because
the D-terms themselves are warped in that model.  This is as opposed to the
present theory, in which the D-terms are equal along the lattice.

We can mediate the SUSY breaking via heavy messengers $T$ and $\overline{T}$
connecting the first lattice site to the MSSM, as in Figure~\ref{fig:king-centipede}.
If we give the messengers a large mass compared to the scale of the light modes
(but potentially $\ll M_{Pl}$), then
the D-term at the first lattice site splits the squared masses of the scalar
fields in $T$ and
$\overline{T}$ by an amount $\pm 2\left<D_1\right>$.  Fermion masses
remain unchanged at tree level.
\EPSFIGURE{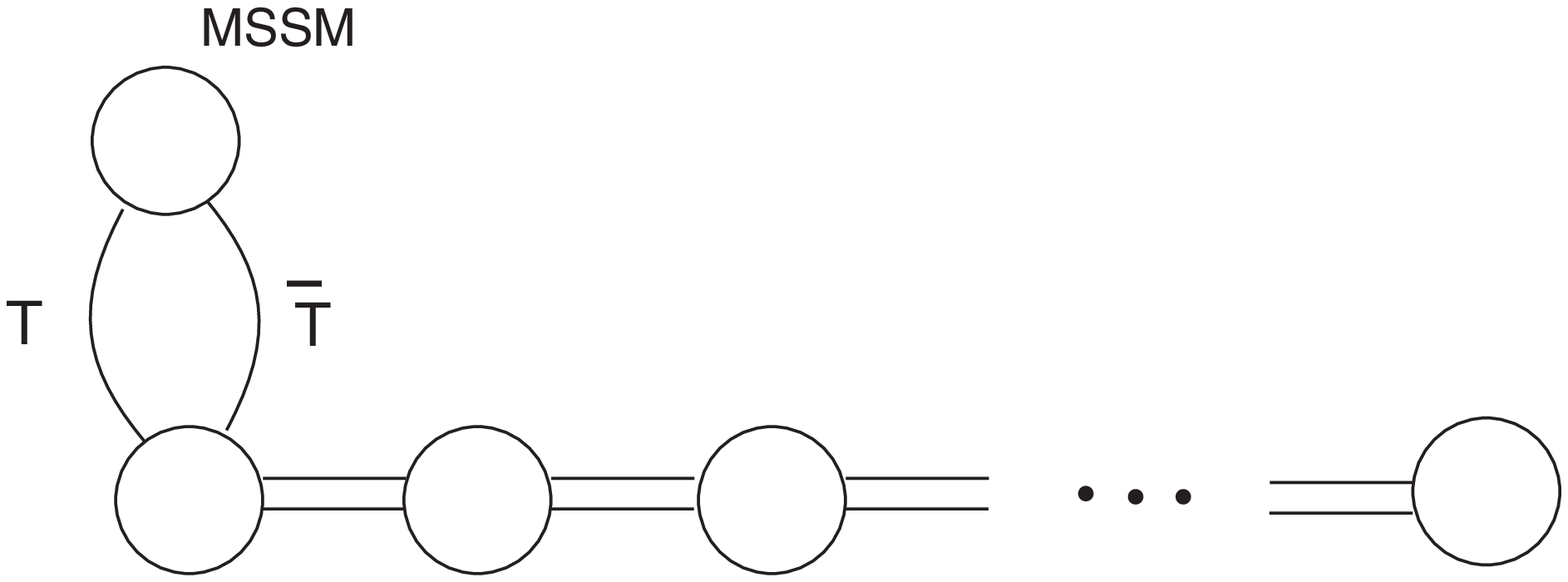,width=4in}{Schematic representation of the coupling of the
U(1)$^n$ theory to the MSSM (or gauge extended MSSM)
via the messenger fields $T$ and
$\bar{T}$. \label{fig:king-centipede}}

Because SUSY breaking is governed by D-terms, the collective D-term breaking
in our model only generates supersoft SUSY breaking terms~\cite{supersoft}.
As exposited in \cite{supersoft}, in the MSSM coupled to a D-term SUSY breaking
sector, SUSY breaking scalar masses first arise by a dimension-ten operator.  If
the gauge sector of the MSSM is enhanced to that of ${\cal N}$=2 SUSY in 4D, then
additional SUSY breaking operators are possible, including supersoft Dirac gaugino masses
$\sim D/M$ and supersoft scalar squared masses $\sim D^2/M^2$.
These operators are called supersoft because they do not give rise to log divergences in
scalar masses or in other operators.  The phenomenology of such models is interesting and,
with the extended gauge sector, the spectrum of the theory interpolates between that of
gaugino mediation and gauge mediation \cite{supersoft}.  However, due to
the collective SUSY breaking in our U(1) moose model, the phenomenology here is somewhat
different.  To examine this, we extend the MSSM gauge sector by adding MSSM adjoint chiral
multiplets $A$, enhancing the gauge sector to the content of ${\cal N}$=2 SUSY multiplets.

F-terms can couple the messengers to the adjoint chiral multiplets of the
extended gauge sector of the MSSM via a Lagrangian of the form,
\begin{equation}
\int d^2\theta\,\left(M T\overline{T}+\alpha T A \overline{T}\right)
\ .\end{equation}
Here we assume that $A$ has vanishing VEVs so that the MSSM gauge symmetry is not
broken by the adjoint scalars.
Then $T$ and $\overline{T}$ do not obtain VEVs, so the U(1) $D$ term
VEVs are unchanged by the presence of the messenger fields.
Integrating out the messengers through diagrams containing the D-term vev
gives rise to supersoft scalar and
gaugino masses \cite{supersoft}.
However, because of the hierarchy in this model between
the vacuum energy and the D-term at each site, the gravitino need not be
the lightest SUSY particle
even if $M\ll M_{Pl}$ with weak scale
supersoft gaugino masses $\sim\left<D_1\right>/M\sim m_W$, since
$m_{3/2}\sim \sum_i \left<D_i\right> / M_{Pl} \gg  \left<D_1\right>/M_{Pl}$ for a
large moose.  This is unlike the generic supersoft theory
in which a single D-term governs the scale of SUSY breaking masses in both
the MSSM and the supergravity sector and the gravitino can be naturally light.

\section{Conclusions}\label{sec:conc}

Deconstruction provides a new paradigm for creating four-dimensional
gauge theories.  At some points in the space of link field vevs, a deconstructed
theory may have a simple description as the latticized version of a gauge
theory in a higher-dimensional space.  At other points, there may
be no simple correspondence, but the theory may nonetheless possess some
interesting phenomenological features of its extra-dimensional cousins.  Much
of the literature has focused on deconstructed flat extra dimensions, in which
all the link fields have equal vevs, while somewhat less attention has been directed
toward warped spaces. In either case, the origin of the link field vevs and the
mechanism that provides naturally for a warping of the theory space have met
little scrutiny.  We have studied a number of explicit models to shed light on
these issues.

In our nonsupersymmetric constructions, we have seen that a combination of translation invariance
in the bulk and boundary corrections to the link field potential are often sufficient
to generate an exponential profile for the link field vevs. In these
examples,  the bulk potential depends only on a few parameters and could be taken general
in form, aside from the constraint of translation invariance. Local minima of the
potential could be found that exhibited the desired warping, without a significant fine-tuning
of parameters.  The physical spectra of gauge and link fields consists of exponential towers,
with a `pseudo-zero mode' for the link tower corresponding to the broken approximate
translational invariance of the moose.  The mass of this state can be raised by including
appropriate higher-dimension operators.  We found that the
first few states in these towers can mimic the results expected for anti-de Sitter space, but
that the spectra overall deviate exponentially from the expected linear dependence on mode number.
Perhaps the most exciting possibility in these models is that this deviation could be discerned
at a future collider.  In this case, one could learn whether the physics observed corresponds
to an underlying theory space or to a new physical space.

In the supersymmetric case, we focused primarily on an Abelian theory, where the warping
was accomplished via Fayet-Iliopoulos D-terms that forced the squares of the link field
vevs to grow by an additive factor as one moves along the moose.  Aside from providing
an existence proof of supersymmetric versions of the type of models of interest to us,
this U(1)$^n$ model is particularly interesting in the way that supersymmetry breaking occurs
non-locally: without all $n$ $D$-flatness constraints (involving all of the $n-1$ link fields)
there would be supersymmetric vacua. In its simplest form, the
model has the peculiar feature that supersymmetry breaking appears only via the generation
of a cosmological constant, while the spectra of the physical gauge and link states remains
supersymmetric.  In the case where the moose is allowed to couple to additional matter, the
delocalization of supersymmetry breaking implies that fields localized at a single site
experience a source of supersymmetry-breaking, $D_i^2$, that is $1/n$ as strong as
the full amount available for gravity mediation leading, for example, to a heavy gravitino.
In addition, supersymmetry breaking is supersoft in this scenario. These features may make
our U(1)$^n$ model distinctive if it is applied as a secluded
supersymmetry-breaking sector for the minimal supersymmetric standard model.

The models we have presented suggest a path for building more realistic and economical
models in warped theory space.  The construction of useful warped non-Abelian supersymmetric
moose models and a study of the full standard model embedding in this framework are directions
worthy of future study.


\section*{Acknowledgments}
C.D.C. thanks the NSF for support under grant PHY-0456525.  J.E. and B.G. thank the NSF
for support under grant PHY-0504442 and the Jeffress Memorial Trust for support under
grant J-768.   J.E. thanks the Aspen Center for Physics,
where some of this work was completed.

\end{document}